\documentclass[conference, 10pt]{IEEEtran}

\newcommand{\design}{T-SAR}

\usepackage{cite}
\usepackage{amsmath,amssymb,amsfonts}
\usepackage{algorithm,algpseudocode}
\algnewcommand\algorithmicforeach{\textbf{for each}}
\algdef{S}[FOR]{ForEach}[1]{\algorithmicforeach\ #1\ \algorithmicdo}
\usepackage{listings}
\usepackage{array}
\usepackage{graphicx}
\usepackage{textcomp}
\usepackage{xcolor}
\usepackage{orcidlink}
\usepackage{svg}
\def\BibTeX{{\rm B\kern-.05em{\sc i\kern-.025em b}\kern-.08em
    T\kern-.1667em\lower.7ex\hbox{E}\kern-.125emX}}
\usepackage{enumitem}
    
\usepackage{multirow}
\usepackage{makecell}
\usepackage{booktabs}
\newcolumntype{P}[1]{>{\centering\arraybackslash}p{#1}}
\newcolumntype{M}[1]{>{\centering\arraybackslash}m{#1}}

\usepackage{siunitx}
\usepackage{pgfplots, pgfplotstable}
\pgfplotsset{compat=1.16}
\usepackage{tikz}
\tikzset{axis background/.style={inner sep=0, outer sep=0}}
\tikzset{every axis label/.style={inner sep=0, outer sep=0}}
\definecolor{color0}{HTML}{1C6E8C}
\definecolor{color1}{HTML}{EE7674}
\pgfplotscreateplotcyclelist{colors}{
{fill=color0},
{fill=color1}
}

\usepackage[font=footnotesize]{caption}
\usepackage{verbatim}
\usepgfplotslibrary{external}
\newcommand\resetstackedplots{
\makeatletter
\pgfplots@stacked@isfirstplottrue
\makeatother
\addplot [forget plot,draw=none] coordinates{(1,0) (2,0) (3,0)};
}

\begin{document}

\title{\fontsize{22pt}{24pt}\selectfont\design: A Full-Stack Co-design for CPU-Only \underline{T}ernary LLM Inference via In-Place \underline{S}IMD \underline{A}LU \underline{R}eorganization}


\author{
\IEEEauthorblockN{Hyunwoo Oh, KyungIn Nam, Rajat Bhattacharjya, Hanning Chen, Tamoghno Das, Sanggeon Yun, \\Suyeon Jang, Andrew Ding, Nikil Dutt, and Mohsen Imani}
\IEEEauthorblockA{
Department of Computer Science, University of California, Irvine \\
Email: \{hyunwooo, m.imani\}@uci.edu}
}

\maketitle

\begin{abstract}

Recent advances in LLMs have outpaced the computational and memory capacities of edge platforms that primarily employ CPUs, thereby challenging efficient and scalable deployment. While ternary quantization enables significant resource savings, existing CPU solutions rely heavily on memory-based lookup tables (LUTs) which limit scalability, and FPGA or GPU accelerators remain impractical for edge use. This paper presents \design{}, the first framework to achieve scalable ternary LLM inference on CPUs by repurposing the SIMD register file for dynamic, in-register LUT generation with minimal hardware modifications. T-SAR eliminates memory bottlenecks and maximizes data-level parallelism, delivering 5.6--24.5$\times$ and 1.1--86.2$\times$ improvements in GEMM latency and GEMV throughput, respectively, with only 3.2\% power and 1.4\% area overheads in SIMD units. T-SAR achieves up to 2.5--4.9$\times$ the energy efficiency of an NVIDIA Jetson AGX Orin, establishing a practical approach for efficient LLM inference on edge platforms. 

\end{abstract}

\vspace{-0.5mm}

\vspace{-1mm}
\section{Introduction}\label{sec:intro}




Large Language Models (LLMs) have become ubiquitous today, with numerous applications, including those in coding assistants, document analysis, and interactive conversational interfaces across consumer and enterprise systems~\cite{llmsurvey}. However, LLMs require substantial resources for inference, with billions of parameters driving extensive matrix operations and creating significant latency during autoregressive generation, where each token requires a full model forward pass~\cite{llmsurvey}.

Traditionally, LLMs have been deployed on cloud servers with high-power GPUs, NPUs, or specialized accelerators to handle their extreme compute and memory demands (e.g., $>$140 GB memory for Llama2-70B~\cite{llama2}). Increasingly, however, there is a need to run LLMs directly on the edge for scenarios such as coding copilots with proprietary source code, document analysis of confidential business data, and personalized assistants handling sensitive information~\cite{edgellm}. In these settings, reliance on cloud computing is often infeasible due to intellectual property concerns, data privacy regulations like GDPR and HIPAA~\cite{gdpr}, limited network connectivity, or prohibitive cloud costs for continuous inference workloads.

Thus, to enable standalone LLM deployment on edge devices at lower cost, several techniques have been proposed, including pruning~\cite{llm_pruner, lazyllm}, quantization~\cite{awq, gpt3int8, onebit, random1}, knowledge distillation~\cite{distill, kdd}, and weight binarization~\cite{binarybert, bibert, pbllm}.

\begin{figure}[tb!]
\centering
\includegraphics[width=0.95\linewidth]{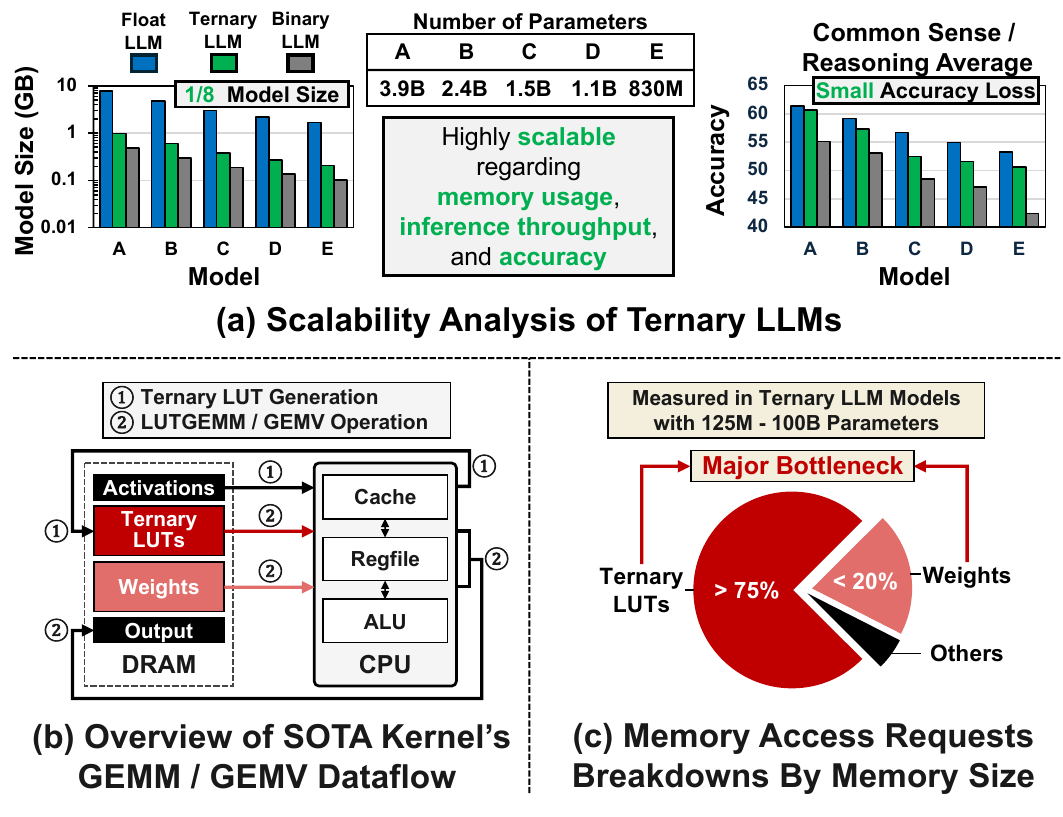}
\vspace{-3mm}
\caption{\textbf{Motivation for scalable ternary LLM acceleration.}
\textbf{(a)} Ternary LLMs provide 8$\times$ size reduction with minimal accuracy loss, making them suitable for edge deployment.
\textbf{(b)} GEMM/GEMV dataflow: SOTA LUT-based kernels store TLUTs in DRAM, causing frequent memory access requests.
\textbf{(c)} Memory access breakdown: TLUTs dominate system memory requests—over 75\%—across models from 125M to 100B parameters, creating a major bottleneck for CPU inference.}
\label{fig:motivation}
\vspace{-5mm}
\end{figure}

Within quantization, ternary quantization has emerged as a particularly promising approach~\cite{bitnet, bitnet_scale, ternary_scale}. By constraining weights to \{-1, 0, 1\}, it achieves 8$\times$ memory compression (Fig.~\ref{fig:motivation}(a)) while maintaining 93–99\% of full-precision accuracy, presenting even better model-size scaling than binary LLMs \cite{ternary_scale}. This enables dramatic cost reduction and practical deployment on resource-constrained edge devices.

However, traditional (edge-) CPUs face a
fundamental architectural misalignment when 
ternary LLMs are deployed, resulting in degraded performance. 
State-of-the-art (SOTA) methods such as T-MAC~\cite{tmac} and BitNet.cpp~\cite{bitnet_cpp} replace multiply-accumulate (MAC) operations with dynamic lookup tables (LUTs) stored in memory (caches and DRAM), shifting workloads from compute-bound to memory-bound ones, achieving 2.4-6.2× the throughput compared to the FP16-based kernels \cite{bitnet_cpp}. As shown in Fig.\ref{fig:motivation}(c), ternary LUT (TLUT) accesses account for over 75\% of system memory requests, creating bandwidth pressure that underutilizes Single-Instruction Multiple-Data (SIMD) execution units ubiquitous in commodity processors~\cite{simd}. This excessive memory traffic cancels out much of the computational benefit of ternary quantization.

While dedicated AI accelerators and FPGAs~\cite{tereffic, tellme, random0} achieve high ternary inference efficiency, their cost and integration complexity preclude widespread edge deployment. Likewise, server-class features such as Intel AMX~\cite{flexinfer} or custom ISA extensions with large compute arrays~\cite{rv5_mpnn, xpulpnn, benini} remain unavailable on edge CPUs due to area and power constraints. In contrast, the SIMD units that current approaches underutilize already provide high-bandwidth register files with datapaths naturally aligned to ternary operations. Yet no prior work has exploited these SIMD registers for in-register LUT computation, leaving an opportunity to overcome memory bottlenecks and fully harness existing parallel hardware.

Therefore, in this paper we introduce \design{}, a full-stack co-design framework for scalable, high-throughput ternary LLM inference on edge CPUs, achieved by leveraging existing SIMD hardware with only minimal modifications. \textbf{The core innovation of \design{} is repurposing the SIMD vector register file for dynamic, in-register LUT generation—eliminating costly memory traffic and maximizing data-level parallelism without new compute arrays or complex datapath extensions.} While we focus on the x86 AVX2 ISA, the idea extends naturally to other SIMD ISAs (e.g., ARM NEON, RISC-V Vector~\cite{neonrvv}), requiring only parameter retuning. \design{}’s design spans four tightly integrated layers:

\begin{itemize}
\item \textbf{Algorithmic Layer:} A ternary-to-binary decomposition and data packing scheme for efficient LUT computing.
\item \textbf{ISA Layer:} Minimal extensions that add register-to-register LUT-based General Matrix Multiplication (GEMM)/General Matrix-Vector Multiplication (GEMV), supporting dynamic computation within SIMD units.
\item \textbf{Microarchitecture Layer:} Power and area overhead analyses based on lightweight wiring/multiplexing adjustments for SIMD units, validated by ASIC synthesis.
\item \textbf{Software Layer:} An adaptive kernel dataflow that maximizes throughput across diverse models and platforms.
\end{itemize}

From high-performance to low-power edge CPUs, \design{} achieves 5.6–24.5$\times$ GEMM latency reduction and 1.1–86.2$\times$ GEMV throughput improvement over SOTA CPU baselines~\cite{tmac, bitnet_cpp} on ternary LLMs from 125M to 100B parameters, with only 3.2\% power and 1.4\% area overhead in SIMD units. Notably, \design{} also delivers 2.5–4.9$\times$ higher energy efficiency than the Jetson AGX Orin GPU on Llama-8B~\cite{bitnet_scale} and Falcon3-10B~\cite{falcon3}.
These results demonstrate that systematic co-design can unlock the latent potential of ubiquitous SIMD hardware for practical edge LLM deployment, narrowing the gap with specialized accelerators while leveraging the world’s most widely deployed compute platform: CPUs.

\section{Motivation}
\label{sec:bg}

We now characterize the bottleneck in SOTA ternary kernels that motivates the co-design framework \design{}.

\begin{figure}[tb!]
\centering
\includegraphics[width=\linewidth]{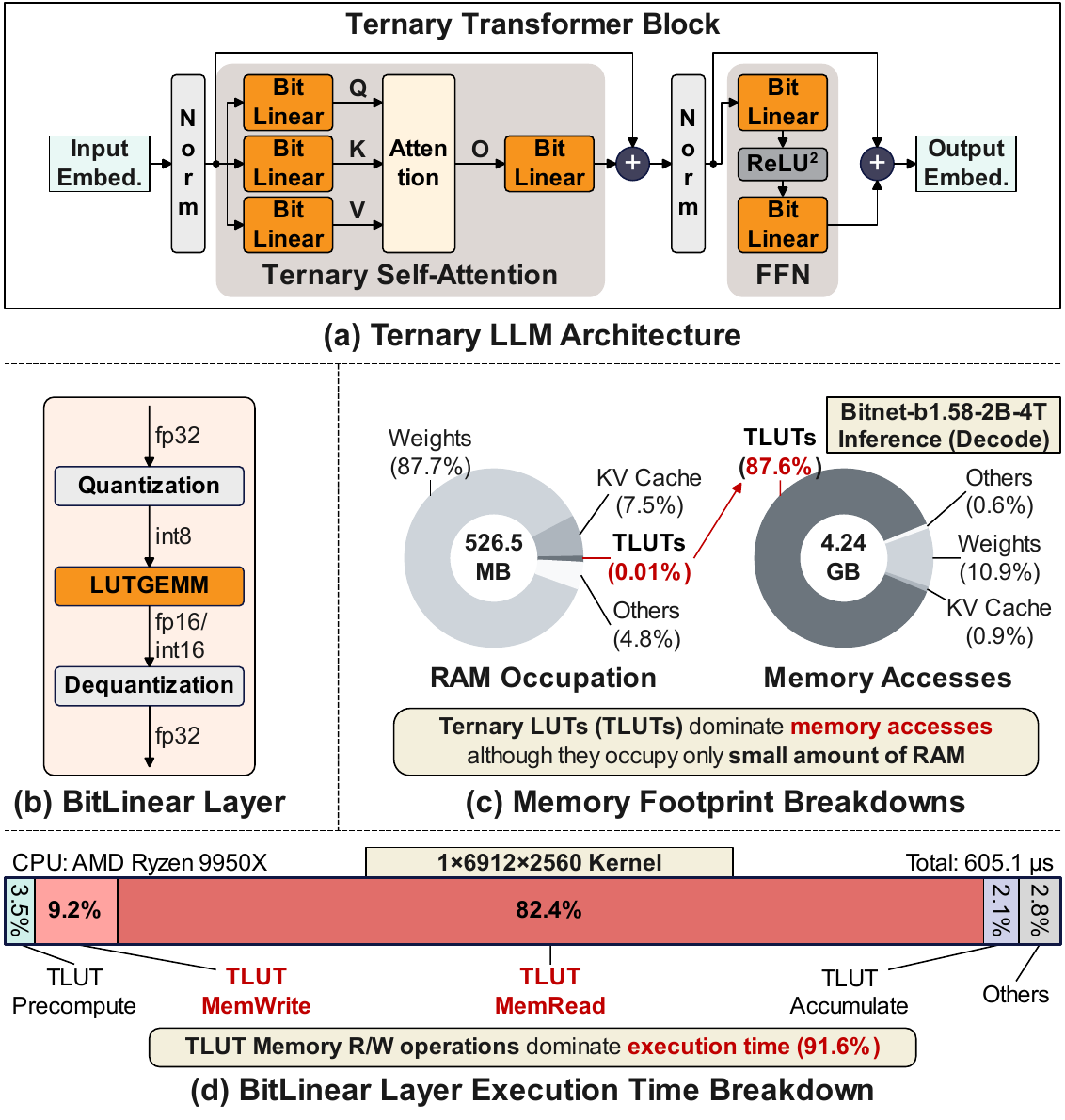}
\vspace{-6mm}
\caption{\textbf{Ternary LLMs: Architecture and bottleneck analysis.}
\textbf{(a)} Ternary transformer with BitLinear layers.
\textbf{(b)} BitLinear layer workflow including quantization and LUTGEMM.
\textbf{(c)} BitNet-b1.58-2B-4T memory footprints: TLUTs, though tiny in RAM, dominate memory accesses.
\textbf{(d)} BitLinear GEMV time breakdown: Memory R/W dominates execution.
}
\label{fig:ternary_llm}
\vspace{-5mm}
\end{figure}

While ternary LLMs promise efficient inference by quantizing weights to $\{-1, 0, 1\}$ and replacing costly floating-point multiplications with low-cost integer operations, the SOTA approach for executing them introduces a critical bottleneck. As shown in Fig.~\ref{fig:ternary_llm}(a,b), these models are implemented in `BitLinear` layers, where the core matrix multiplication is handled by an \textbf{LUT-based GEMM/GEMV} method~\cite{tmac, bitnet_cpp}. 
This technique partitions each input vector (with $K$ total inputs) into atomic blocks of size $c$. For each block, all $3^c$ possible dot product results are pre-computed and stored in an LUT, 
transforming the computation from $O(K)$ arithmetic operations per output to $O(K/c)$ table lookups. 
The total LUT storage per layer becomes $O((K/c) \cdot 3^c)$, 
that trades off
arithmetic complexity for memory accesses,
resulting in the
fastest CPU implementations today.

\begin{figure}[tb!]
\centering
\includegraphics[width=\linewidth]{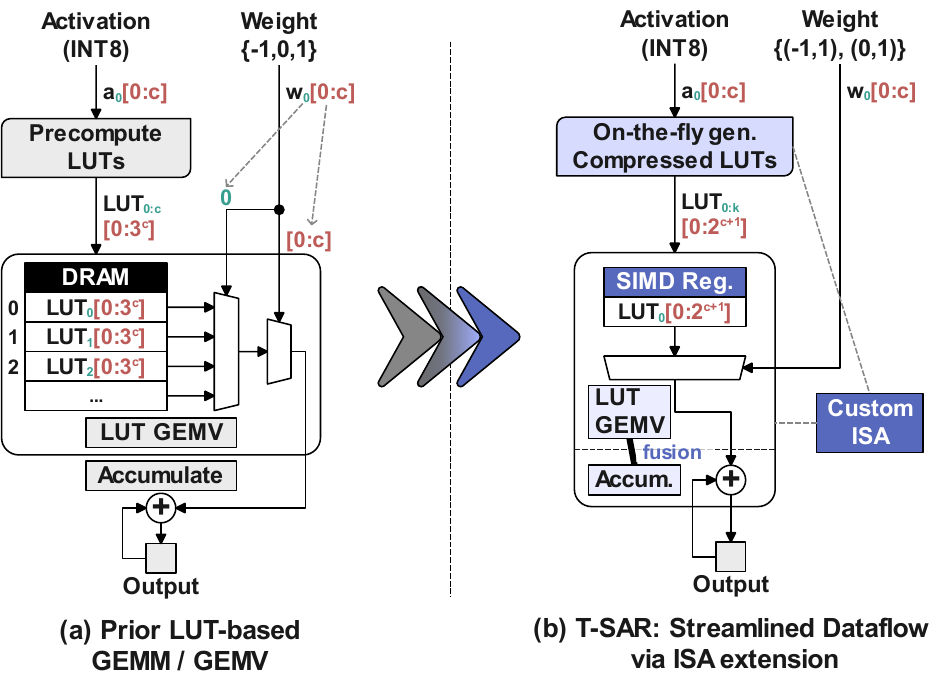}
\vspace{-6mm}
\caption{\textbf{Prior LUT-based CPU solution vs. \design{}.}
(a) Prior: precomputed LUTs loaded from DRAM.
(b) T-SAR: on-the-fly compressed LUTs generated in SIMD registers.
}
\label{fig:intro_diff}
\vspace{-6mm}
\end{figure}

Though theoretically efficient, this trade-off introduces a critical \textbf{memory access bottleneck}. As Fig.~\ref{fig:intro_diff}(a) illustrates, the SOTA dataflow relies on pre-computing all LUT values and storing them in system memory. During inference, weights are used to index these LUTs, requiring frequent, random memory accesses that fail to efficiently utilize modern cache hierarchies and powerful SIMD hardware. Our analysis reveals the severity of this issue. Although ternary LUTs (TLUTs) occupy less than 0.01\% of total RAM in a representative model, they are accessed so frequently that they account for a staggering \textbf{87.6\% of all memory transactions} (Fig.~\ref{fig:ternary_llm}(c)). This memory saturation means the workload becomes \textbf{memory-bound rather than compute-bound}, with \textbf{91.6\% of the execution time} spent on memory read/write (R/W) operations (Fig.~\ref{fig:ternary_llm}(d)).

To overcome these limitations, \design{} aims to \textbf{eliminate external LUT loads and fully exploit the SIMD datapath}, shown in Fig.~\ref{fig:intro_diff}(b). \design{} generates compressed LUTs on-the-fly inside SIMD registers via custom ISA extensions, eliminating memory TLUT traffic and supporting fused GEMV-accumulation operations to increase throughput. However, achieving in-register LUT computation is non-trivial: the required LUT size ($3^c$) does not match the fixed $2^n$ bitwidth of SIMD registers, and the limited register file size constrains the number of LUTs that are held simultaneously. This necessitates the careful co-design detailed in the following sections.

\section{\design{} Co-Design Stack}
\label{sec:stack}
To resolve the memory bottleneck mentioned in Section~\ref{sec:bg}, \design{} introduces a full-stack co-design that transforms LUT-based operations from 
memory-bound to compute-bound tasks.
The core innovation is 
that we
eliminate memory traffic by generating LUTs on-the-fly, directly within the CPU's high-speed SIMD register file. This section details the tightly integrated algorithmic, ISA, and microarchitectural layers that enable this transformation.

\subsection{Algorithm: Ternary-to-Binary Decomposition}
The primary challenge of in-register LUT generation is the architectural mismatch between the base-3 nature of ternary weights and the base-2 structure of SIMD hardware. A naive LUT for a block of $c$ weights would require $3^c$ entries, which does not align with SIMD register bitwidths.

\begin{figure}[tb!]
\centering
\includegraphics[width=\linewidth]{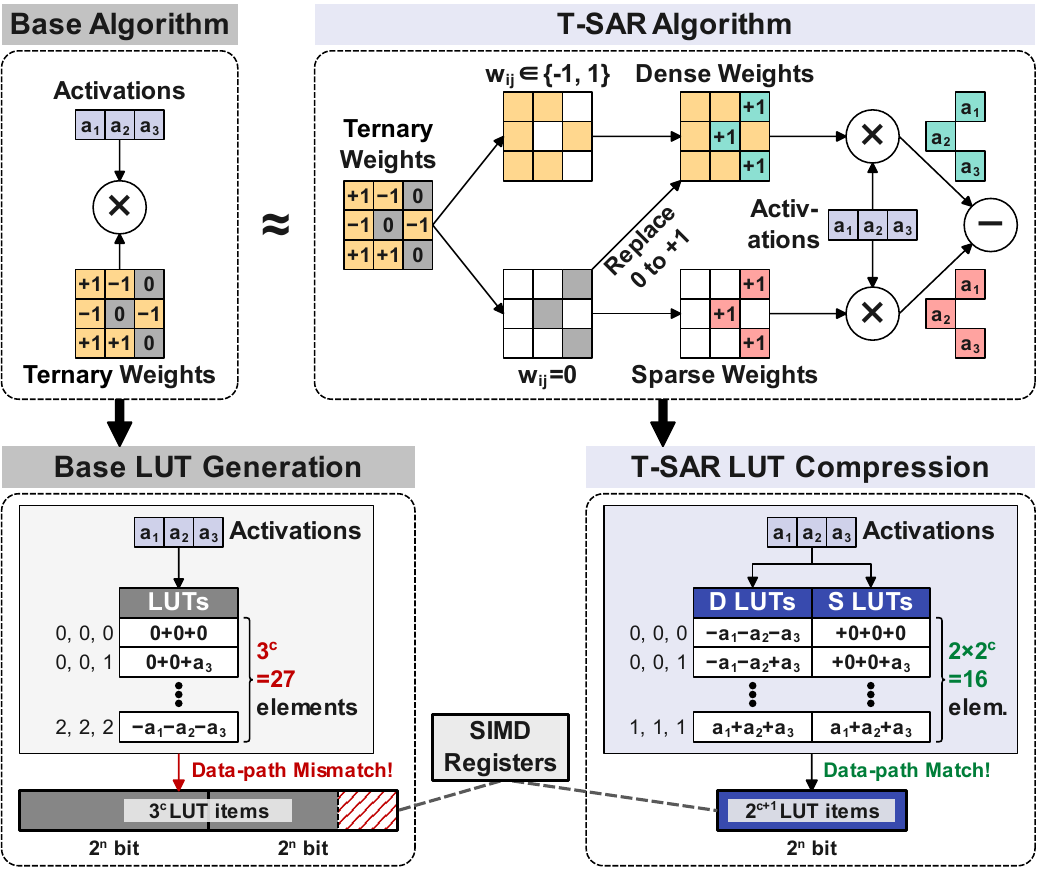}
\vspace{-5mm}
\caption{\textbf{Proposed LUT GEMV Algorithm for LUT compression, matching the LUT size to the data-path.}}
\vspace{-2mm}
\label{fig:lut_compression}
\end{figure}

\design{} overcomes this with a novel \textbf{LUT compression via weight transformation}, as shown in Fig.~\ref{fig:lut_compression}. We decompose a ternary weight block $\mathbf{w} \in \{-1,0,1\}^c$ and its corresponding input activations $\mathbf{a} \in \mathbb{R}^c$ into two separate binary forms:
\begin{itemize}
    \item \emph{Dense weights}: $\mathbf{w}_D \in \{-1, +1\}^c$, where $w_{D,i} = w_i$ if $w_i \neq 0$, else $+1$.
    \item \emph{Sparse weights}: $\mathbf{w}_S \in \{0, 1\}^c$, where $w_{S,i} = 1$ if $w_i = 0$, else $0$.
\end{itemize}
This allows the original dot product to be re-expressed as a subtraction of two binary dot products, removing the influence of the zero-weighted elements:
\[
y = \sum_{i=1}^c w_i a_i = \sum_{i=1}^c w_{D,i} a_i - \sum_{i=1}^c w_{S,i} a_i
\]
With this transformation, instead of one ternary LUT with $3^c$ entries, we only need two binary LUTs (for $\mathbf{w}_D$ and $\mathbf{w}_S$), each of size $2^c$. The total storage per block becomes $2^{c+1}$, which perfectly matches the power-of-two width of SIMD registers and avoids significant data-path augmentation.

\begin{figure}[tb!]
\centering
\includegraphics[width=\linewidth]{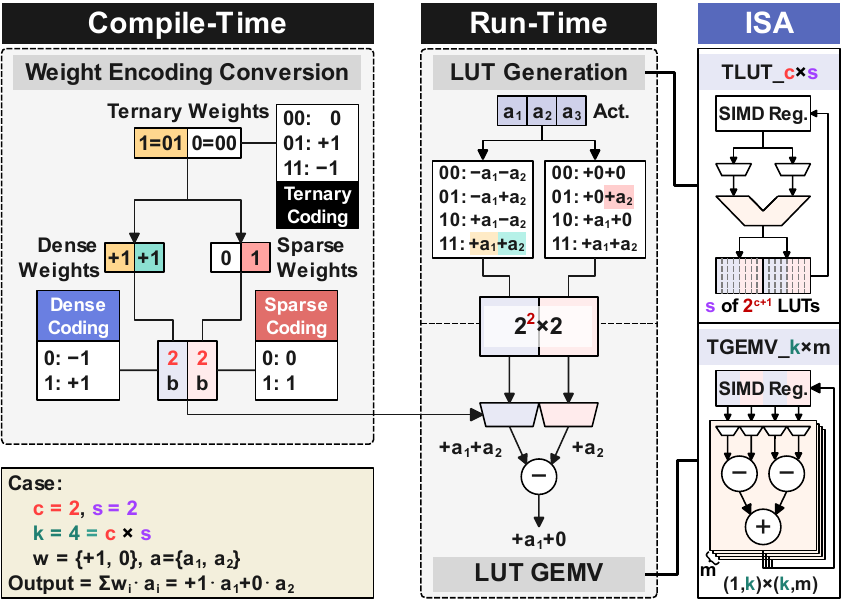}
\vspace{-5mm}
\caption{\textbf{\design{}'s LUT-based kernel framework overview.}}
\vspace{-3mm}
\label{fig:computation}
\end{figure}

\begin{figure*}[tb!]
\centering
\includegraphics[width=\linewidth]{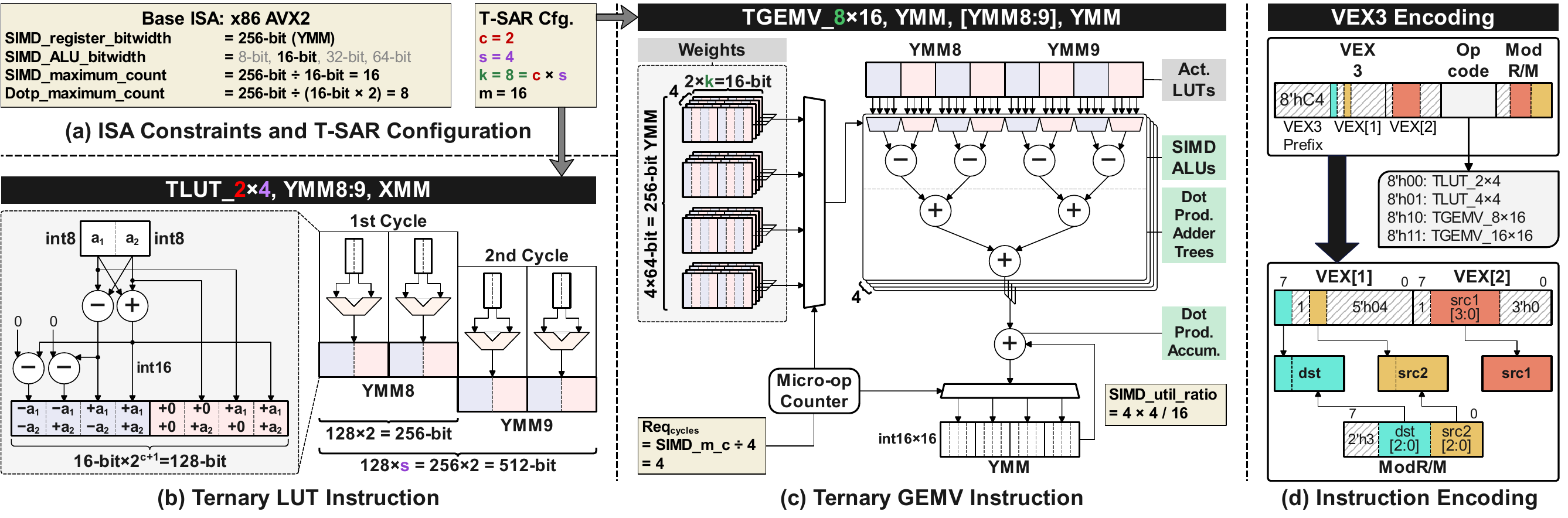}
\vspace{-5mm}
\caption{\footnotesize{\textbf{\design{}'s ISA extension applied to x86 AVX2 SIMD ISA}, demonstrating how the \design{} instruction primitives are realized with only minimal hardware changes, utilizing existing SIMD ALUs and adder trees. \textbf{(a)} \design{} configuration and base ISA constraints. \textbf{(b)} \texttt{TLUT\_c×s} example. \textbf{(c)} \texttt{TGEMV\_k×m} example. \textbf{(d)} Instruction encoding details for AVX2, with designed examples of \texttt{TLUT\_c×s} and \texttt{TGEMV\_k×m} instructions.}}
\vspace{-3mm}
\label{fig:isa}
\end{figure*}

\subsection{ISA Extensions for In-Register Execution}
As shown in Fig.~\ref{fig:computation}, the decomposition mentioned above enables a two-phase kernel framework. At compile time, ternary weights are encoded into dense and sparse binary forms. At run time, a minimal set of ISA extensions execute the LUT-based GEMV directly within SIMD registers. These instructions are parameterized by:
\begin{itemize}
    \item $c$: the block size.
    \item $s$: the number of input blocks processed per instruction.
    \item $k$: the number of input channels processed per instruction ($k = c \times s$).
    \item $m$: the number of output channels.
\end{itemize}
The workflow is: (1) the \texttt{TLUT\_c×s} instruction generates two binary LUTs from activations and places them in SIMD registers; (2) the \texttt{TGEMV\_k×m} uses these register-resident LUTs with pre-encoded weights to perform the $(1,k)\!\times\!(k,m)$ GEMV; and (3) the final ternary result is reconstructed by subtraction. This register-resident design aligns computation with the SIMD datapath, eliminating the memory bottleneck.

\subsection{Microarchitectural Implementation}
Fig.~\ref{fig:isa} demonstrates a practical ISA extension realization on the x86 AVX2 ISA. For the example configuration shown in Fig.~\ref{fig:isa}(a), we set $c\!=\!2$, $s\!=\!4$, $k\!=\!8$, and $m\!=\!16$.

The \texttt{TLUT\_2×4} instruction (Fig.~\ref{fig:isa}(b)) generates four register-resident LUTs, each with $2^{c+1}\!=\!8$ 16-bit entries, occupying two 256-bit YMM registers\cite{ymm} ($4\!\times\!8\!\times\!16\!=\!512$ bits). To minimize hardware changes, the operation is split into two $\mu$-ops, each writing 256 bits per cycle.

The \texttt{TGEMV\_8×16} instruction (Fig.~\ref{fig:isa}(c)) performs a $(1,8)\!\times\!(8,16)$ GEMV, producing 16 outputs. It involves $s\!\times\!m\!=\!64$ subtractions and $m\!=\!16$ $s$-to-1 adder tree (ADT) operations, distributed over four $\mu$-ops. These instructions reuse the existing 256-bit YMM datapath, including all 16 16-bit SIMD ALUs and 4-to-1 ADTs originally for dot product instructions. Only minor wiring and multiplexer additions are required, keeping area and power overhead minimal.

The instruction encoding, detailed in Fig.~\ref{fig:isa}(d), uses standard VEX3 fields for both TLUT and TGEMV primitives. For instructions that span multiple registers (e.g., \texttt{TLUT\_2×4} writing to YMM8:9 or \texttt{TGEMV\_8×16} reading from YMM8:9), the destination or source register is interpreted as a register pair: if dst is 0x1000, the operation uses YMM8 and YMM9.

\vspace{-0.5mm}
\subsection{Software Kernel and Dataflow Optimization}
\begin{figure}[tb!]
\centering
\includegraphics[width=\linewidth]{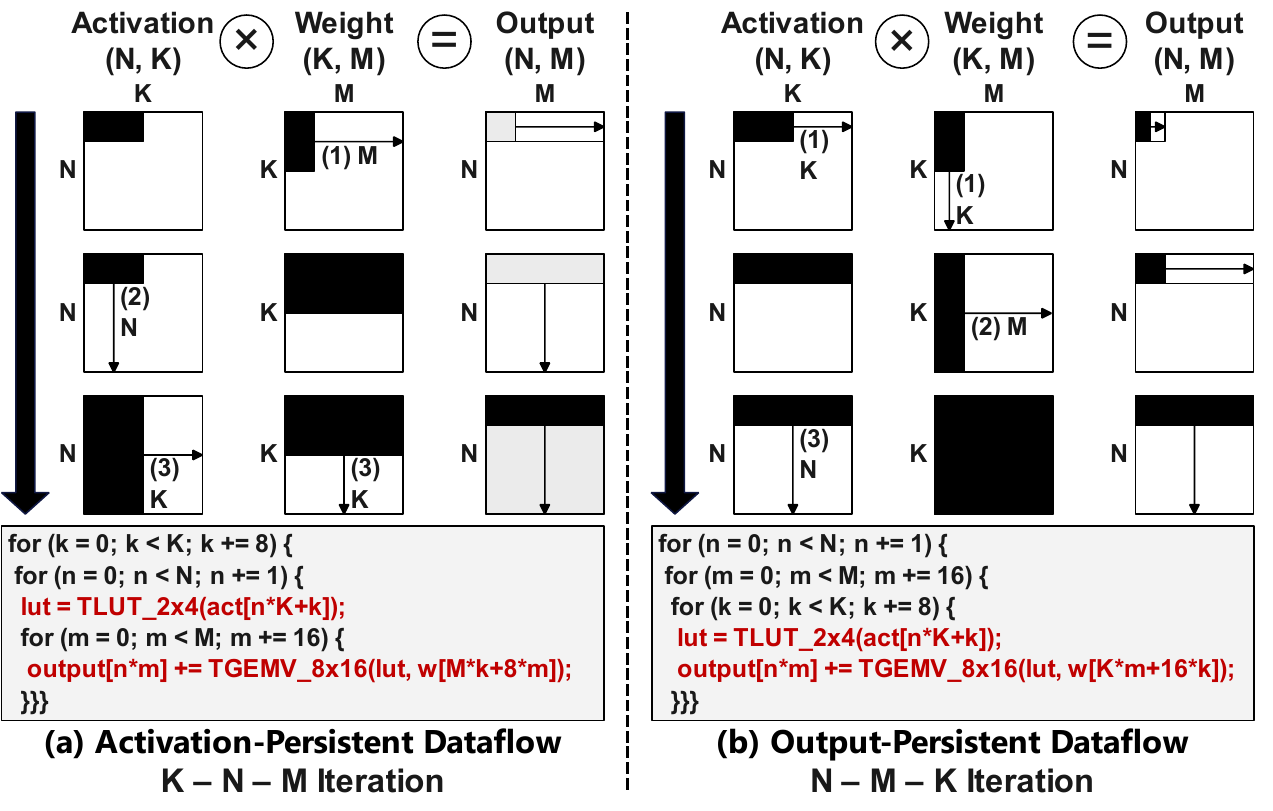}
\vspace{-5mm}
\caption{\textbf{The \design{}'s kernel dataflow selections.} \textbf{(a)} Activation-persistent dataflow minimizing the \texttt{TLUT\_c×s} invocations and increases input/weight cache hits. \textbf{(b)} Output-persistent dataflow reducing total memory footprints.}
\vspace{-3mm}
\label{fig:kernel}
\end{figure}

The performance of GEMM/GEMV operations is highly dependent on data reuse patterns, which vary across different layers of an LLM. To maximize throughput, \design{} employs an adaptive kernel scheduling strategy. We implement two microkernel dataflows—\emph{activation-persistent} (AP) and \emph{output-persistent} (OP)—to flexibly match these patterns (Fig.~\ref{fig:kernel}).

The AP dataflow (Fig.~\ref{fig:kernel}(a)) retains input activations in registers across the inner loop, minimizing LUT recomputation and increasing input and weight cache hits. This is effective for layers with high activation and weight reuse (i.e., high $N$ and $K$). In contrast, the OP dataflow (Fig.~\ref{fig:kernel}(b)) keeps output accumulators local until computation completes, reducing memory write-back traffic. This is beneficial in layers with a high number of output channels (high $M$). \\
\noindent
At compile-time, \design{}’s inference framework empirically selects the fastest kernel for each layer, ensuring maximum performance across the entire model.

\begin{table*}[t]
\centering
\caption{gem5 simulator configurations for evaluation platforms.}
\vspace{-2mm}
\label{tab:system_cfg}
\resizebox{\textwidth}{!}{%
\begin{tabular}{c|cccccccc}
\toprule
System Type             & CPU Model             & Simulation Mode               & Cores & Freq. & L1 I/D Cache     & L2 Cache          & L3 Cache      & DRAM          \\ 
\midrule\midrule
\textbf{Workstation}    & AMD Ryzen 9950X       & \multirow{3}{*}{DerivO3CPU}   & 16 & 5.7\,GHz & 32\,KB / 48\,KB  & 1\,MB/core        & 64\,MB shared & DDR5-6400\,MHz \\
\textbf{Laptop}         & AMD Ryzen 7840U       &                               & 8  & 5.1\,GHz & 32\,KB / 32\,KB  & 1\,MB/core        & 16\,MB shared & DDR5-4400\,MHz \\
\textbf{Mobile}         & Intel Processor N250  &                               & 4  & 3.8\,GHz & 64\,KB / 32\,KB  & 2\,MB shared      & 6\,MB shared  & DDR5-4400\,MHz \\
\bottomrule
\end{tabular}}
\vspace{-4mm}
\end{table*}

\section{Experiments and Evaluation}\label{sec:evaluation}  
We now evaluate \design{} across diverse models and platforms to validate three key claims: (1) \design{} delivers significant end-to-end speedups for both prefill (GEMM-heavy) and decode (GEMV-heavy) phases in autoregressive LLMs; (2) these improvements arise from fundamentally reducing the memory bottleneck of SOTA LUT-based methods; and (3) they are achieved with minimal hardware overhead, making \design{} highly efficient even compared to edge GPUs.

\subsection{Experimental Setup}

\textbf{ISA and Simulator:}
We extend \texttt{gem5-AVX} 20.1.0.0 \cite{gem5, gem5-avx} (DerivO3CPU) to model the \design{} ISA.  New \texttt{TLUT\_c×s} and \texttt{TGEMV\_k×m} operations are added to the AVX2 pipeline with cycle‑accurate $\mu$‑op sequencing and register‑pair reads/writes. Each instruction was verified by executing hand‑written assembly with byte‑pattern encodings.

\textbf{Kernels:}
We design three kernel variants for each LUT-GEMV pair (\texttt{TLUT\_2$\times$4}+\texttt{TGEMV\_8$\times$16}, \texttt{TLUT\_4$\times$4}+\texttt{TGEMV\_16$\times$16}), resulting in six kernels:
(1) AP-min: activation-persistent with minimal register usage;
(2) AP-max: activation-persistent with maximal register use to reduce iterations;
(3) OP: output-persistent for minimal write-back traffic.
All are implemented in C++ with inline assembly and compiled with GCC 9.4.0.

\textbf{Baselines:}
We compare against two SOTA LUT-based baselines: Bitnet.cpp TL-2~\cite{bitnet_cpp} and T-MAC~\cite{tmac}.
To ensure fairness, our kernels include both input quantization and output dequantization stages (shown in Fig.~\ref{fig:ternary_llm}(b)).

\textbf{Models and Protocol:}
We evaluate BitNet models from 125M to 100B parameters~\cite{bitnet_infra}.
Prefill runs with $N\!=\!128$ tokens (batch=1) to build the KV cache; decode measures steady-state throughput using the KV cache.
Thread counts are fixed at \{16, 8, 4\} for \{Workstation, Laptop, Mobile\}.

\textbf{Metrics:}
We report prefill latency, decode throughput, kernel execution time, and kernel memory access requests.

\textbf{Platforms:}
Three representative x86 CPU classes are modeled (Table~\ref{tab:system_cfg}):  
\textbf{Workstation} (Ryzen 9950X, 16 cores),  
\textbf{Laptop} (Ryzen 7840U, 8 cores),  
\textbf{Mobile} (Intel N250, 4 cores).

\subsection{End-to-End Results}
\begin{figure}[tb!]
\centering
\vspace{-2mm}
\includegraphics[width=\linewidth]{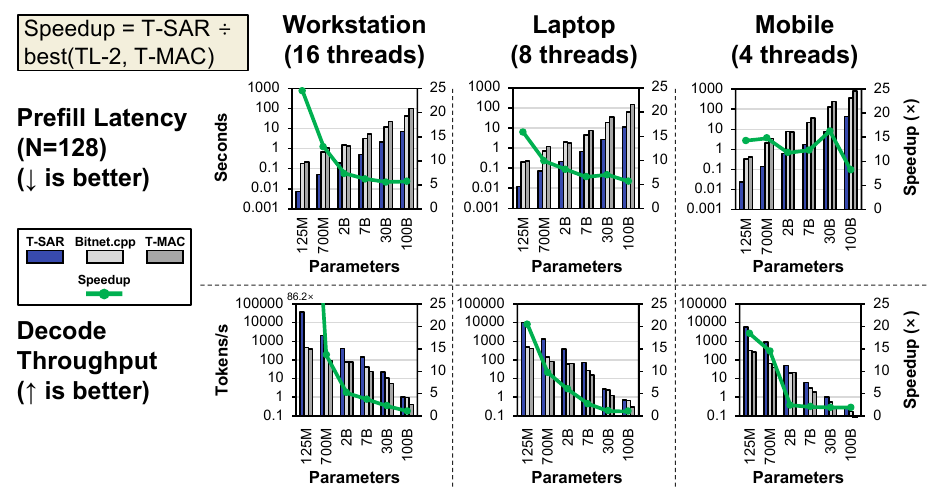}
\vspace{-6mm}
\caption{\textbf{End-to-end performance across platforms.}}
\vspace{-3mm}
\label{fig:graph_e2e}
\end{figure}
\textbf{Prefill (GEMM-heavy):}
As shown in Fig.~\ref{fig:graph_e2e} (top), \design{} delivers geo-mean prefill speedups of 8.8× (Workstation), 8.4× (Laptop), and 12.4× (Mobile) across all models (125M–100B).  
Since GEMM is compute-bound, register-resident LUT generation and fused accumulation let efficiency gains translate almost directly into throughput until cache/memory contention emerges.  
This explains why prefill benefits exceed decode gains.  
For example, Mobile’s 7B prefill drops from $>$20\,s to under 1.7\,s—enabling interactive LLM use on devices where GPUs cannot be deployed.

\textbf{Decode (GEMV-heavy):}
Fig.~\ref{fig:graph_e2e} (bottom) shows that baseline GEMV is dominated by repeated TLUT fetches. \design{} removes this traffic entirely, exposing SIMD compute throughput.  
Relative gains peak on Workstation (6.4×), due to larger caches delaying memory-system bandwidth saturation; Laptop and Mobile reach 4.1–4.2×.  
Mobile’s smaller gain reflects earlier bandwidth saturation despite the request-volume reduction.

\textbf{Link to bottlenecks:}
The prefill/decode gap mirrors GEMM’s compute-bound and GEMV’s bandwidth-bound nature—setting up the trends seen in memory-system analysis.

\vspace{-1mm}
\subsection{Memory-System Impact}
\begin{figure}[tb!]
\centering
\vspace{-3mm}
\includegraphics[width=\linewidth]{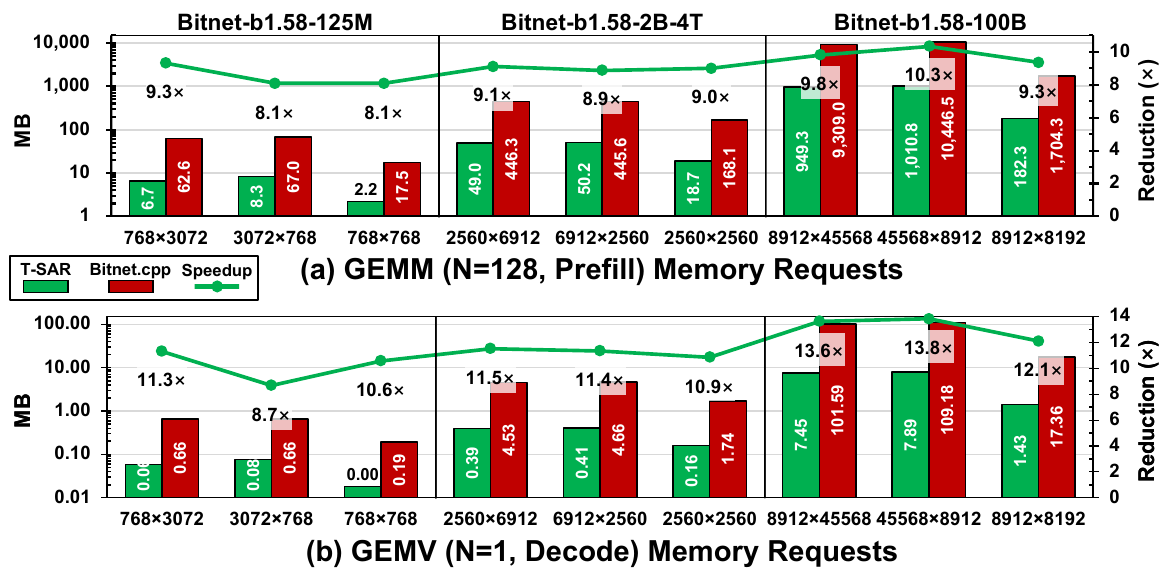}
\vspace{-7mm}
\caption{\textbf{Memory request volume (MB) of the kernels executed in the representative BitNet model inference (125M, 2B-4T, 100B)}.  
(a) GEMM (N=128) prefill. (b) GEMV (N=1) decode.}
\vspace{-5mm}
\label{fig:graph_mem_req}
\end{figure}

The central claim of \design{} is that it removes the memory bottleneck by generating LUTs directly in registers. As shown in Fig.~\ref{fig:graph_mem_req}, this reduces memory request volume (MB) by 8.7–13.8× compared to TL-2~\cite{bitnet_cpp}, with GEMV showing larger relative cuts because the baseline is TLUT-dominated and TLUT fetches are eliminated. For GEMM, TL-2’s denser weight packing (1.67\,bits/weight) limits relative reductions, but the resulting stall decreases still improve ALU utilization.

Request reduction grows with model size ($K\!\times\!M$) as more LUT calls are avoided, but latency gains diverge:  
(1) \textbf{GEMV case-} Large cuts yield smaller returns as lower Last-level cache (LLC) hit rate reduces effective bandwidth and forces early saturation—most evident in Mobile 1$\times$8192$\times$45568: 89\%$\rightarrow$62\%—capping latency drops.  
(2) \textbf{GEMM case-} Even modest cuts yield large drops since compute-bound phases convert freed cycles and higher locality into utilization; LLC hit rate stays high (89\%$\rightarrow$91\%) until contention.  
These effects predict the thread-scaling behavior shown in Fig.~\ref{fig:graph_kernel_latency}.

\vspace{-1mm}
\subsection{Kernel Microbenchmarks and Thread Scaling}
\begin{figure*}[tb!]
\centering
\includegraphics[width=\linewidth]{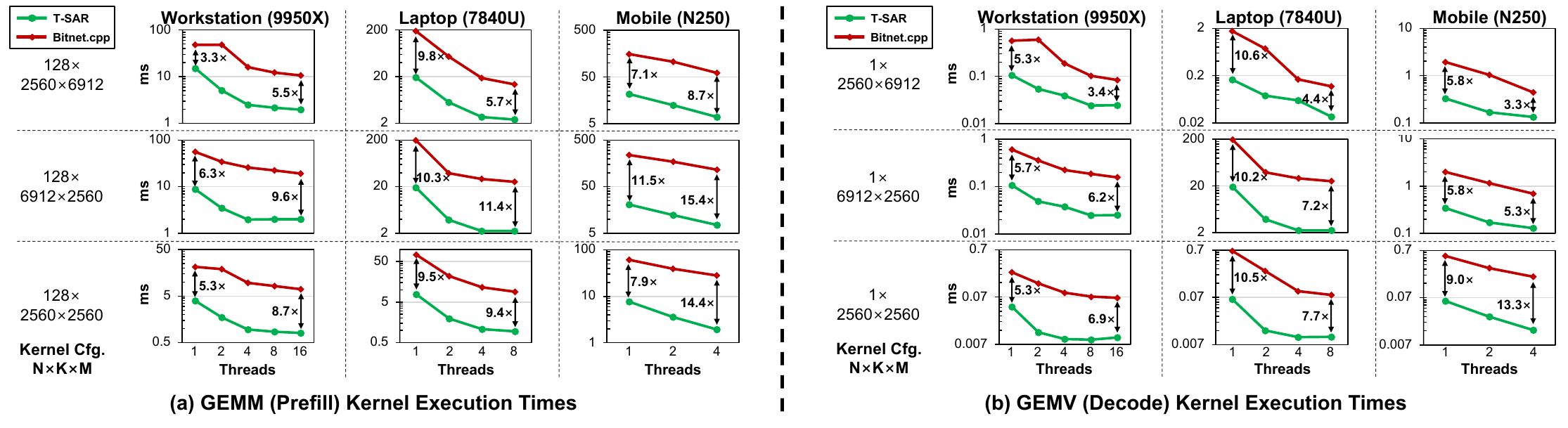}
\vspace{-6mm}
\caption{\textbf{Multi-thread scaling for BitNet-b1.58-2B-4T. \design{} vs.\ TL-2 for GEMM (left) and GEMV (right).}  
Solid lines denote absolute latency (log scale). Arrows show relative speedup.}
\vspace{-5mm}
\label{fig:graph_kernel_latency}
\end{figure*}
From Fig.~\ref{fig:graph_kernel_latency}, we make the following observations:

\textbf{GEMM case:}
For large shapes (128×2560×6912, 128×6912×2560), \design{} sustains scaling up to 8–16 threads (Workstation) and 4–8 (Laptop) before L3/DRAM contention dominates.
Scaling is not perfectly linear, but flattens later than GEMV due to the compute-bound nature and higher data locality, allowing freed cycles from reduced stalls to be fully exploited.
This yields up to $13\times$ speedup at 4 threads—matching the large prefill gains in Fig.~\ref{fig:graph_e2e}.

\textbf{GEMV case:}
Despite the largest proportional memory savings, GEMV saturates effective bandwidth quickly—often by 2–4 threads on Mobile and 4–8 on Workstation/Laptop—leading to early plateaus and smaller decode-time improvements.  
Here, extra cores cannot offset bandwidth limits.

Hence, we see that compute-bound kernels (GEMM) sustain scaling and achieve larger absolute gains across platforms, while memory-bound kernels (GEMV) plateau quickly—highlighting \design{}’s impact and limits for real-world deployment for edge platforms.\footnote{{\underline{Note:}}
We report prefill with $N{=}128$ due to simulation cost. TL-2’s weight packing (1.67-bit) is denser than our 1+1-bit split, resulting in approximately 20\% more static memory occupation, but end-to-end is dominated by TLUT traffic rather than weight RAM size, hence \design{}’s advantage. We demonstrate our framework to x86 due to the broad applicability, but retargeting to NEON or RISC-V Vector (RVV)~\cite{neonrvv} only requires $c,s,k,m$ tuning due to the different SIMD lane width but extant dot product extensions. For instance, existing \textbf{ARM NEON's 128-bit datapath} with SDOT/UDOT instruction support (since ARMv8.2-A \cite{neon_dotp}) \textbf{realizes the \texttt{TLUT\_2×4} + \texttt{TGEMV\_8×8}}.
}

\vspace{-0.5mm}
\subsection{Hardware Overheads}
To size the cost of \design{}, we synthesized a 256-bit SIMD unit (vector add/mul/dot-product, write-back interface) at 1\,GHz in TSMC 28\,nm using Cadence Genus~21.10, both \emph{without} and \emph{with} the \design{} logic (shown in Table~\ref{tab:hw_overhead}). Multi-mode multi-corner (MMMC) synthesis covered \emph{ssg$\leftrightarrow$ffg}, $V_{\!DD}\!\in[0.81,1.05]$\,V, $T\!\in[0,125]^\circ$C; Area and power are reported at \texttt{tt0p9v25c}. The \design{} instructions reuse existing ALU lanes and register file—no new arithmetic units or scratchpads. Additions are: (i) a 256-bit vector \emph{write-back} MUX to inject TLUT words into the register file, (ii) small operand-bus wires and input MUXes for TLUT/TGEMV (no extra read ports), and (iii) a tiny control/scoreboard block to sequence TLUT writes and fused accumulation. 


\begin{table}[h]
\centering
\caption{Synthesis of a 256-bit SIMD slice (TSMC 28\,nm, 1\,GHz) with and without \design{} ISA.}
\vspace{-2mm}
\label{tab:hw_overhead}
\setlength{\tabcolsep}{6pt}
\renewcommand{\arraystretch}{1.06}
\resizebox{\linewidth}{!}{%
\begin{tabular}{lrrrrrr}
\toprule
\multirow{2}{*}{Block} &
\multicolumn{3}{c}{Area ($\mu$m$^2$)} &
\multicolumn{3}{c}{Power (mW)} \\
\cmidrule(lr){2-4}\cmidrule(lr){5-7}
& Base & \design{} & $\Delta$ & Base & \design{} & $\Delta$ \\
\midrule
SIMD ALUs + write-back interface        & 73{,}560 & 73{,}560 & \,\,0.0\% & 5{,}904 & 5{,}904 & \,\,0.0\% \\
\design{} $\rightarrow$ write-back MUX  & 0        & \,\,588  & \textbf{+0.8\%} & 0     & \,\,41  & \textbf{+0.7\%} \\
Operand-bus wires and input MUX         & 0        & \,\,147  & \textbf{+0.2\%} & 0     & \,\,24  & \textbf{+0.4\%} \\
Others (control/scoreboard, decode)     & 0        & \,\,295  & \textbf{+0.4\%} & 0     & 121     & \textbf{+2.0\%} \\
\midrule
\textbf{Total}                              & \textbf{73{,}560} & \textbf{74{,}590} & \textbf{+1.4\%} &
                                              \textbf{5{,}904} & \textbf{6{,}090} & \textbf{+3.2\%} \\
\bottomrule
\end{tabular}}
\vspace{-2mm}
\end{table}

\textbf{Result:} As shown in \autoref{tab:hw_overhead}, the area increases by \textbf{+1.4\%} (73{,}560$\to$74{,}590\,$\mu$m$^2$) and active power consumption under kernel-like switching rises by \textbf{+3.2\%} (5{,}904$\to$6{,}090\,mW), dominated by toggling on the new MUX paths. Results correspond to a single 256-bit SIMD slice (no SRAM arrays) and exclude caches and register files; absolute area will scale with the number of slices and integration.

\subsection{Cross-Platform Comparison}
To further evaluate our CPU-based solution, we compare it directly against an edge GPU in the NVIDIA Jetson AGX Orin~\cite{orin} SoC, using identical model checkpoints and runtime settings (batch=1; steady-state decode). 
For CPUs,  \design{} power is estimated from CPU \emph{package} power under TL-2 decode via the measured dynamic overhead in Table.~\ref{tab:hw_overhead}:
$P_{\text{\design{}}} = 1.032 \cdot P_{\text{TL-2}}$.
Energy per token is $E = P_{\text{\design{}}}/(\text{tokens/s})$ from the measured \design{} throughput from gem5 simulator.

\begin{table}[h]
\centering
\caption{Cross-platform decode throughput and energy/token (batch=1). Power boundary: CPU package; GPU module.}
\vspace{-2mm}
\label{tab:cross_combined}
\setlength{\tabcolsep}{5pt}
\renewcommand{\arraystretch}{1.04}
\small
\resizebox{\linewidth}{!}{%
\begin{tabular}{lcc|cc}
\toprule
& \multicolumn{2}{c}{\textbf{Llama-b1.58-8B}} & \multicolumn{2}{c}{\textbf{Falcon3-b1.58-10B}}\\
\cmidrule(lr){2-3}\cmidrule(lr){4-5}
\textbf{Platform} & tokens\,/\,s & J\,/\,token & tokens\,/\,s & J\,/\,token \\
\midrule
Workstation CPU (9950X, 4nm, \design{}) & 128.96 & 0.616 & 103.93   & 0.795 \\
Laptop CPU (7840U, 4nm, \design{})      &  61.00 & 0.405 & 49.65    & 0.540 \\
Mobile  CPU (N250, 10nm, \design{})     &   5.18 & 0.733 & 4.30     & 0.953 \\
\midrule
Jetson AGX Orin GPU (8nm, llama.cpp)    &  16.78 & 1.839 & 13.25    & 2.620 \\
\bottomrule
\end{tabular}}
\vspace{-2mm}
\end{table}

\textbf{Takeaways.}
With matched per-model checkpoints and settings, \design{} on CPUs outperforms Jetson on \textbf{Workstation} and \textbf{Laptop} across both families:
Llama-b1.58-8B shows 7.7× / 3.0× (tokens\,/\,s / lower J\,/\,token) on workstation and 3.6× / 4.5× on laptop;
Falcon3-b1.58-10B shows 7.8× / 3.3× and 3.7× / 4.9×, respectively.
On \textbf{Mobile}, throughput is lower than Jetson (0.31–0.32×), yet energy/token remains 2.5–2.75× lower, consistent with our decode memory-bandwidth analysis.

\section{Conclusion}\label{sec:conclusion}  
We presented \design{}, a full-stack co-design framework for CPU-only ternary LLM inference. The key idea is to move LUT generation from memory into SIMD registers, turning TLUT fetches into in-register compute and fusing accumulation so BitLinear layers shift from bandwidth-bound to datapath-bound execution. This yields portable speedups across CPUs with no new ALUs and only minor mux/control logic, supported by AP/OP kernels that adapt per layer to each model and platform. Beyond the reported gains, the approach generalizes to RVV and NEON~\cite{neonrvv}, invites advanced LUT compression and compiler scheduling, and integrates naturally with sparsity or further quantization. In short, a small ISA extension realigns the bottleneck and enables interactive LLM inference on CPUs—even mobile ones—while keeping hardware changes minimal.

\bibliographystyle{unsrt}
\bibliography{References/intro}

\end{document}